\documentclass[12pt,letterpaper]{article}
\usepackage{graphicx}
\usepackage{sagetex}
\usepackage{amssymb}
\usepackage{caption}
\usepackage{local.defs}
\begin{document}

%\ttfamily
%\slshape
\rmfamily
\begin{center}
\vspace*{1in}
\Ls {\bf Matrix Characterization of Knots:\\
     A Simple Statistical Mechanics Application}\ns

\vspace{1in}

                R. Kariotis   \\
         {\em Department of Physics \\
        University of Wisconsin \\
          Madison, Wisconsin 53706} \\
       (~/text/p.fn.tex --  28jun13) \\
\end{center}

Abstract: \\

In this note, I describe a formalism for treating knots as geometric
spaces, and make an application to a simple statistical mechanics
computation. 
The motivation for this study is the natural visual symmetry of the knot,
and I describe how this might be carried out. The direct approach,
however, fails due to limits of the visual symmetry, but
by recasting the problem in terms of the geometry of contours of
the knot, the resulting permutation operators provide a better
analytical tool. At present, I will be limited to a
description of the proposed program, rather than complete results.

In the process of setting up the formalism, an interesting
aspect, the use of fractional permutation operators, turned up,
and I will present a simple statistical mechanical application
of these objects.
The application described suggests a relation
to {\em anyons}, the key element in topological quantum computing,
which in turn is related to the association of a knot  - in its
braid form - with a quantum circuit.

\np
\noi
{\bf Introduction}

Knot theory [1,2] has become an important area of investigation in theoretical
physics in recent years due to applications in such diverse areas as
quantum gravity, polymer physics and quantum computation [3,4].
(By far the best introduction to this subject is Kauffman's book
{\em Knots and Physics} [5].)
One important
element in the theory which has yet to reach maturity might be described
as the lack of
a clear analytic means of characterizing knots, in particular, of being
able to unambiguously distinguish one knot from another.
Polynomial invariants offer a means for dealing with this need, but are
not unique; the Burau and Artin representations are insufficiently
descriptive. (However, it should be mentioned that computer graphics
 have been used (HFK theory[8])).
The incident of the "Perko Pair" [6] is a fair indication of
the complications involved.

In this note I will outline a possible means for dealing with this need
by setting up a formalism that associates each path/contour 
 around a knot with
a point in an abstract space, $\mathfrak{C}$, the space of contours.
Successive contours are related by a displacement operator, and the set
of such operators then describes a geometry on $\mathfrak{C}$.
Interpolating between points in $\mathfrak{C}$ suggests assigning
metrical meaning to the calculation of fractional permutation
operators. I will make a special example of this by applying
these matrices to a simple statistical mechanics problem.
\\
\\
\noi
{\bf Motivation}

The obvious visual symmetry in the simplest knots
suggests an underlying mathematical symmetry, 
for example as is apparent in the 3.1 knot of the figure below.
A single arc, $\zeta$, can be used to generate the entire
diagram by applying appropriate displacements
(labeled $\gamma$) and rotations (labeled $\Gamma$).
Graphics software that has the appropriate underlying
algebraic facilities, such as {\em METAPOST}, can be
very effective in carrying out these operations and
has the additional advantage  of emphasizing the geometric nature
of the diagram [9a,b].

This approach, however, quickly becomes impractical
as the knots increase in complexity. In the next figures, the
9.2 and 7.2 knots are shown with suggested basis arcs, but
additional pieces would be needed to complete the job,
simple as these diagrams are.

Thus, we follow a different  geometric approach described in
the next section. As it happens, this method is not unlike
the geometric approach once considered by Polish and British
cryptographers during WWII.
In their scheme, a line of encoded text was treated as a
set of permutations of the original text, so the operation
becomes one of a displacement in the space of text strings,
from one line to another. In the context of knots, one contour
along the knot is related to another by a permutation matrix,
a displacement in the space of contours.
\\
\\
\noi
{\bf Geometry}

Starting at a particular point, a contour is defined as the sequence
of numbered crossings, a positive integer for cross-over, negative
for cross-under. An example is shown in figure for the knot 9.2 [7].
The starting point for the contour is arbitrary so
what we'll want is the complete set of numbered contours.

There are two lines of interest here:
\\
\noi
 1) by investigating
the set of permutation operators $\{\gamma_{i}\}$ in the space
$\mathfrak{C}$ of contours, it is possible
to obtain a generalized set of skein relations, constructed
by matrix operations on the space $\{\gamma_{i}\}$;
\\
\noi
 2) it is
useful also to make a geometric interpretation of the space
of operators, defining displacement, angle and curvature;
in this view, the odd-numbered knots, 3.1, 5.1, 7.1, ... are
equivalent to a Euclidean geometry, where the distance from
one contour to the next is uniform along a trajectory in
$\mathfrak{C}$;
all other knots
are non-uniform and it will be useful to take the radical of
products - e.g. $\sqrt{\gamma_{1}\gamma_{2}}$, which sometimes results
in a complex operator - in order to calculate curvature.

\begin{equation}
\begin{array}{rcccccccccl}

 10.161 &     &      &      &     &         &     &      &      &      & 10.162 \\
        &     &      &      &     &         &     &      &      &      &        \\
   \zla &     &      &      &     &         &     &      &      &      & \zra   \\
        &\sea &      &      &     &         &     &      &      & \swa &        \\
        &     & \dla &      &     &         &     &      & \dra &      &        \\
        &\nea &      & \sea &     &         &     &\swa  &      & \nwa &        \\
   \zla &     &      &      & \lR &  U_{12} & \rR &      &      &      & \zra   \\
        &\sea &      & \nea &     &         &     &\nwa  &      & \swa &        \\
        &     & \dla &      &     &         &     &      & \dra &      &        \\
        &\nea &      &      &     &         &     &      &      & \nwa &        \\
   \zla &     &      &      &     &         &     &      &      &      & \zra   \\
\end{array}
\end{equation}
\noi
For example, I show here a scheme where
the set of operators relating the two Perko knots[6]; these were
originally called 10.161 and 10.162 in the Rolfson scheme until Ken Perko
recognized that they were the same knot.

In general, the relation between two successive displacements, and their
composite is given by
\begin{equation}
\begin{array}{ccc}
   \zeta_{1}\gamma_{1}=\zeta_{2}     &  
  \zeta_{1}\Gamma_{1}=\bar{\zeta_{1}}    & 
 \bar{\zeta_{1}}\bar{\gamma_{1}}=\bar{\zeta_{2}}  \\
\end{array}
\end{equation}
\\
\\
\noi
{\em Euclidean metric:} In this case, displacement, defined generally
\[
                   \zeta_{i}\gamma_{i}=\zeta_{i+1}
\]
so each operator acts to connect one contour  to the next,
is such that all $\gamma_{i}$ are equal; in all other knots
this operator varies from one point in the function
space to the next. The diagram here shows the 3.1 knot,
indicating how one might go about constructing the the full
image from a single arc that can be rotated and displaced in
successive moves. Also, below that is shown the 9.2  knot,
with indications of which arcs would be needed to construct a full
image; in fact, considerable effort was needed in the 7.2 knot
and the result is not entirely convincing

In other words, the visual symmetry is a good indication of the
basic topology, but is not very practical as a computational device.

\begin{figure}[!h]
\begin{center}
\includegraphics[width=3in,height=3in]{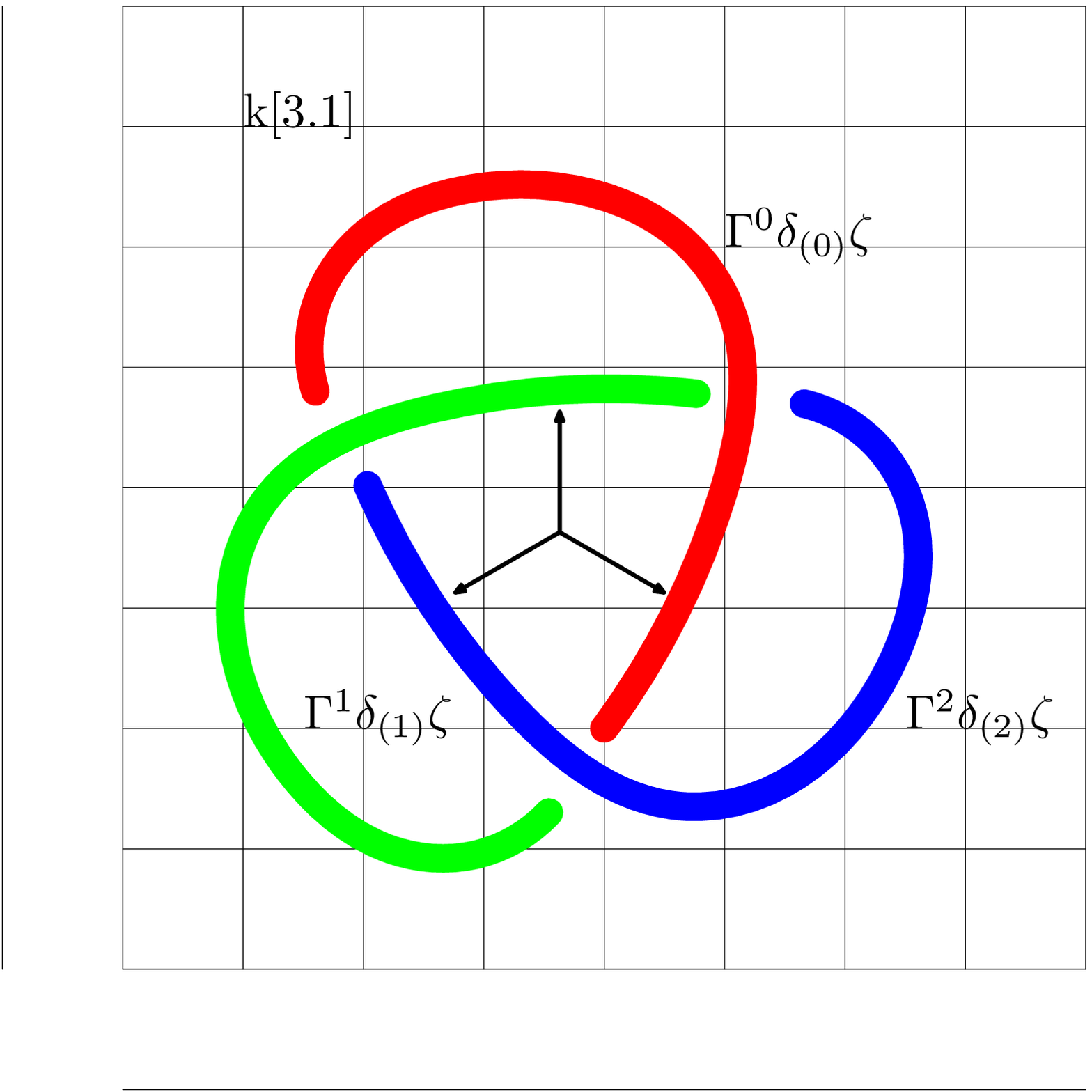}
\caption*{[3.1] arc $\zeta$ displaced by $\gamma$, rotated by $\Gamma$}
\end{center}
\end{figure}

\begin{figure}[!h]
\begin{center}
\includegraphics[width=3in,height=3in]{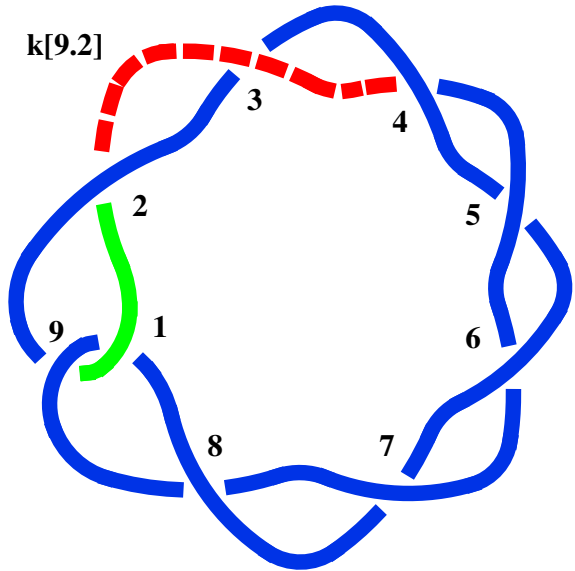}
\caption*{[9.2] Numbering of one contour/path for Rolfson 9.2}
\end{center}
\end{figure}

\begin{figure}[!h]
\begin{center}
\includegraphics[width=3in,height=3in]{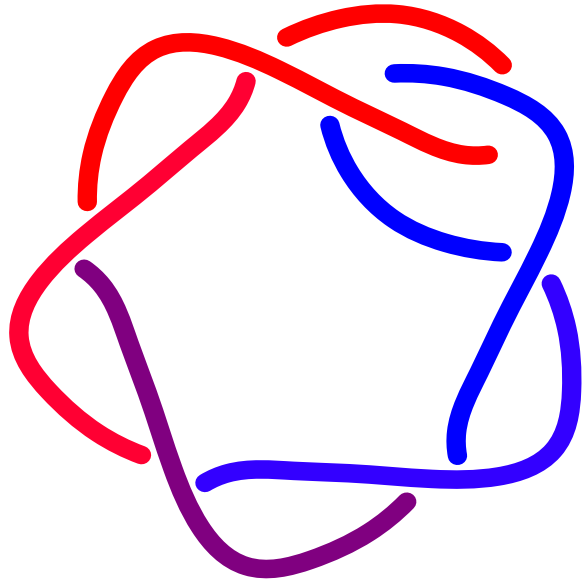}
\caption*{[7.2] construction  using two arcs}
\end{center}
\end{figure}
\begin{figure}[!h]
\begin{center}
\includegraphics[width=3in,height=3in]{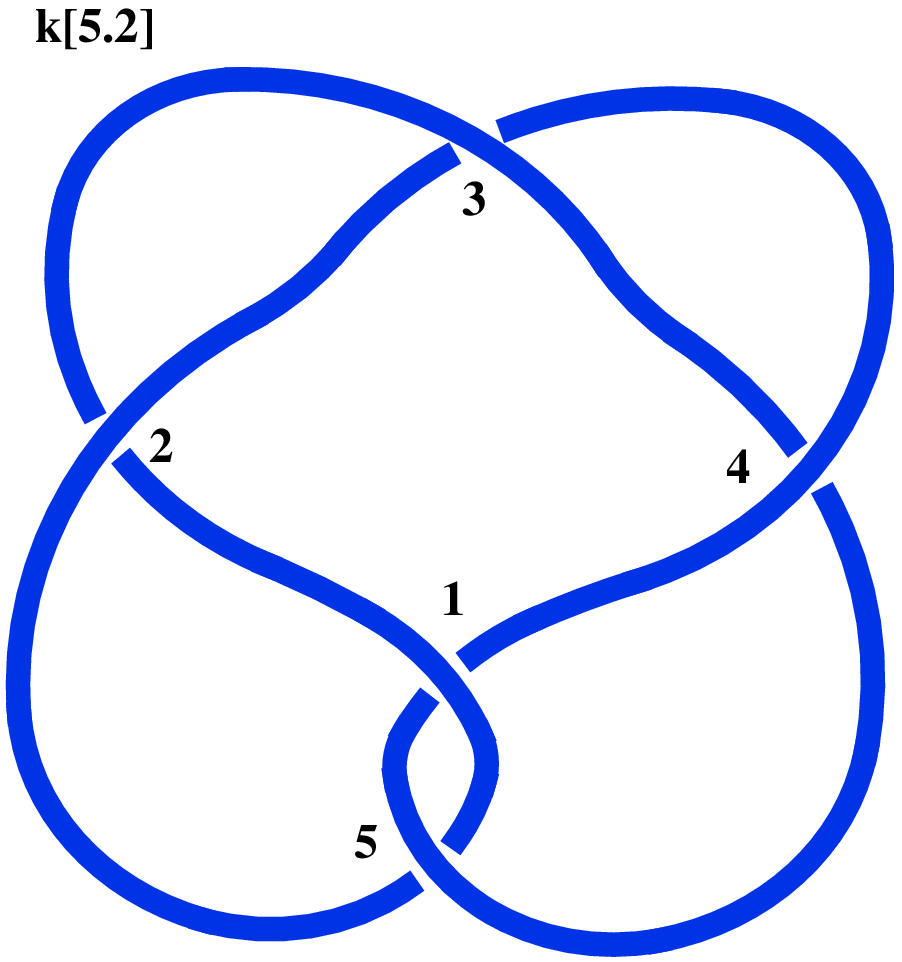}
\caption*{[5.2] typical contour numbering}
\end{center}
\end{figure}
\noi
{\em Non-Euclidean metric:}
The 5.2 diagram shown below, and the set of paths around its contour
suggest the topology of the knot: the operation that carries one
path into the next varies as one moves along a trajectory in
contour space $\mathfrak{C}$.
A measure of the non-uniformity is obtained by considering
the extent to which products of operators can be represented
by a product of identical operators.
 This requires that we be
able to take, for example
\[
                          \sqrt{\gamma_{1}\gamma_{2}}=Q_{1}
\]
in general, however, the resulting radical is complex with multiple roots.
\\
\noi
{\em Curvature:}
In order to determine what we mean by curvature, consider first
the knots with constant $\gamma_{i}$, i.e. 3.1, 5.1, 7.1, ... .
In this case, the displacement of two next near neighbors,
defined as $\gamma_{i}\gamma_{i+1}$ is simply $\gamma_{i}^{2}$ so the
space has uniform metrical properties. All other knots
are non-uniform and the separation of next near neighbors
must be defined as $\sqrt{\gamma_{i}\gamma_{i+1}}$, which in general
is complex.
\np
\begin{equation}
\begin{array}{ccccccccccc}
\zeta_{1}= & \{ 1& -2&   3&  -4&   5&  -1&   4&  -3&   2&  -5 \}\\
\zeta_{2}= & \{ 5& -1&   2&  -3&   4&  -5&   3&  -2&   1&  -4 \}\\
\zeta_{3}= & \{ 4& -5&   1&  -2&   3&  -4&   2&  -1&   5&  -3 \}\\
\zeta_{4}= & \{ 3& -5&   4&  -1&   2&  -3&   1&  -4&   5&  -2 \}\\
\zeta_{5}= & \{ 2& -5&   4&  -3&   1&  -2&   3&  -4&   5&  -1 \}\\
\zeta_{6}= & \{ 1& -4&   3&  -2&   5&  -1&   2&  -3&   4&  -5 \}\\
\zeta_{7}= & \{ 5& -3&   2&  -1&   4&  -5&   1&  -2&   3&  -4 \}\\
\zeta_{8}= & \{ 4& -2&   1&  -5&   3&  -4&   5&  -1&   2&  -3 \}\\
\zeta_{9}= & \{ 3& -1&   4&  -5&   2&  -3&   5&  -4&   1&  -2 \}\\
\zeta_{0}= & \{ 2& -3&   4&  -5&   1&  -2&   5&  -4&   3&  -1 \}\\
\end{array}
\end{equation}

\setlength{\unitlength}{1cm}
% \linethickness{.4cm}
\begin{picture}(10, 7)(-1, -1)
\thicklines
           \put(6.5,2.5){ $\gamma_{2}$ }
           \put(2.5,2.7){$\gamma_{1}$}
           \put(4.7,2){${\hat \rho}$}
           \put(2,0.35){$\sqrt{\gamma_{1}\gamma_{2}}$}
           \put(6.5,0.35){$\sqrt{\gamma_{1}\gamma_{2}}$}
           \put(4,-.1){${\hat \rho}=\exp[{\hat n}]$}

  \put(5,1) {\vector(1,4){.8}}
  \put(1,1) {\vector(3,2){4.7}}
  \put(5.85,4.1) {\vector(1,-1){3.1}}

  \put(1,1) {\vector(1,0){4}}
  \put(5,1) {\vector(1,0){4}}
\put(3,-1) {curvature: normal vector ${\hat n}$}
\end{picture}
\\
\\
\noi
{\em Roots of Permutation Operators}
To see how this comes about, first note that permutation matrices can
always be put into the form of a set of cycles, for example,
$(561432)\rightarrow(153)(26)(4)$ so that by rearranging the indices
the matrix is put in block form
\[
             A=
\left(
\begin{array}{cccc}
                    A_{1} &      &       &       \\
                          & A_{2}&       &       \\
                          &      & A_{3} &       \\
                          &      &       & \ddots\\
\end{array}
\right)
\]
and each block has the form (taking the 4x4 for example)
\begin{sagesilent}
A=load('m')
\end{sagesilent}
\[
       A_{j}=    \sage{A}
\]
(or, for nxn $(n12345...n-1)$)
and matrices of this form are easily diagonalized; a
typical eigenvector is
\[
                     \psi = \left(
\begin{array}{c}
                              \lm^{1}\\
                              \lm^{2} \\
                              \lm^{3} \\
                              \vdots  \\
                              \lm^{n} \\
\end{array}
                             \right)
\]
where the $m^{th}$ eigenvalue is
\[ 
               \lm_{m}=e^{\phi_{m}} \hspace*{1in}
                     \phi_{m}=m\frac{2\pi i}{n}
\]
 For example, the three element matrix is
\[
\left(
\begin{array}{ccc}
                 0 & 1 & 0\\
                 0 & 0 & 1\\
                 1 & 0 & 0\\
\end{array}
\right) 
\rightarrow
\left(
\begin{array}{ccc}
                 e^{\phi}  & e^{2\phi} & e^{3\phi} \\
                 e^{2\phi} & e^{4\phi} & e^{6\phi} \\
                 e^{3\phi} & e^{6\phi} & e^{9\phi} \\
\end{array}
\right)
 \hspace*{.5in}
                     \phi=\frac{i2\pi }{3}
\]
in general, for integral $Y$, there will be $Y$ roots for each of the
 diagonal elements of
$A^{\frac{1}{Y}}$, and which one to apply must be dealt with in context.
If $Y$ is rational, the number of roots is finite; if it is irrational
the number of roots is infinite, resulting in the so called {\em differential
permutation group}, far outside the scope of this work.
For example: if $Y=11$, there will be $11$ roots, located at 
$\phi=n\frac{2\pi}{11}$ where $1\leq n \leq11$;
for $Y=11.1$ there are $111$ roots at
$\phi=10 n\frac{2\pi}{111}$ and $1\leq n \leq111$;
for $Y=11.11$,
$\phi=100 n \frac{2\pi}{1111}$ with $1\leq n \leq1111$.
And so on, as suggested in the diagram.
More on this in the next section where a statistical mechanics application
is described (see also the Appendix).

\begin{figure}[!h]
\begin{center}
\includegraphics[width=1.5in,height=1.5in]{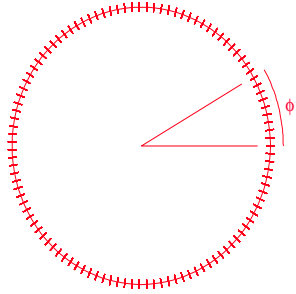}
\caption*{k=11.1 incommensurate frequencies}
\end{center}
\end{figure}

One interesting consequence of this is that the square root induces
an algebra on the group of operators that make the cyclic group.
For example
\[
      S_{1}^{\frac{1}{2}}=\mathfrak{B}S_{1}
                         +\mathfrak{D}S_{1}^{2}
                         +\mathfrak{F}S_{1}^{3}
                         +\mathfrak{H}S_{1}^{4}
                         +\mathfrak{J}S_{1}^{5}
\]
or taking
\[ S_{1}=C_{5}=
\sage{A}
\]
and generally, all roots are sums of the group elements
\[
       T^{\frac{1}{Y}}=
           \aleph \sage{A}
          +\beth \sage{A*A}
          +\daleth \sage{A*A*A}
          +\cdots
\]
where $\aleph, \beth, \daleth, ...$ are numerical constants,
which will simplify the computation considerably. The coefficients
are readily determined from the eigenvalues and eigenvectors; for
example
\[
            \aleph=\lm_{1}\sqrt{\lm_{1}}\lm_{1}^{2}
            +\lm_{2}\sqrt{\lm_{2}}\lm_{2}^{2}
            +\lm_{3}\sqrt{\lm_{3}}\lm_{3}^{2}
\]
where for $C_{3}$, $\lm_{1}=e^{\phi}$ and $\phi=\frac{i2\pi}{3}$.
The only complication is that there are $2^{N}$ roots that must
be considered.

 This formalism is somewhat similar to the versors  of geometric
algebra [10] where the $(\frac{1}{Y})^{th}$ root of a rotation operator
is most easily handled in Clifford algebra formalism.
\\
\\
\\
\noi
{\bf Spin Net Representation}

Each displacement operator
$\Gamma_{i}$ can be expressed as a direct sum of cycles as described
above
\[
             \Gamma_{i}= C_{1} \oplus  C_{2} \oplus  C_{3} \oplus \cdots
\]
unless the geometry of the knot is simple (as in 3.1, 5.1, 7.1, ...),
and each cycle acts as an independent particle state.
In going from one displacement to the next the independent cycles are
transformed: for example in the image below, spin net diagrams 
the first set of cycles $\{C_{i}\}$, on the left, is transformed to
the out going set $\{\bar{C}_{i}\}$.

A diagram of this behavior is shown in Fig. A, and suggests the structure
sometimes found in spin nets [11]. What is lacking at this point is the
operation that takes
\[
                 C_{1}\oplus C_{2}\oplus C_{3}\rightarrow 
                            \bar{C}_{1}\oplus \bar{C}_{2}
\]
That is, that we need is something like that described by Dorst et al in their book
{\em Geometric Algebra and Computer Science} [10], incrementally
transforming $C \rightarrow \bar{C}$. In his example [sect.1.2.6],
a rotation $V$, expressed in versors, is implemented in $N$ steps of
$V^{\frac{1}{N}}$.
\begin{figure}[!h]
\begin{center}
\includegraphics[width=4in,height=3in]{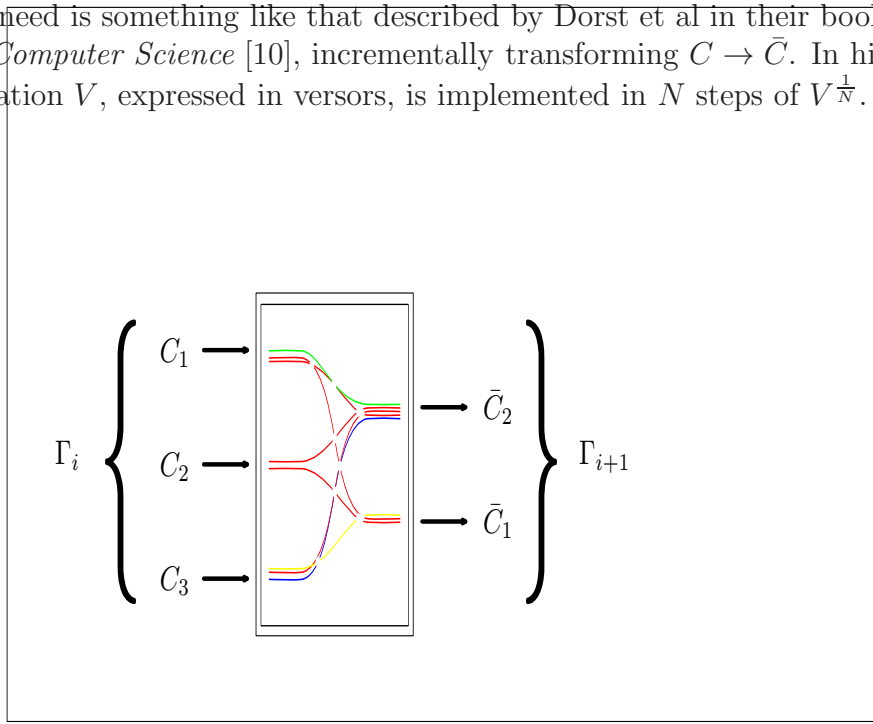}
\caption*{Fig. A: braid image}
\end{center}
\end{figure}
\\
\\
\noi
{\bf Partition Function/Path Integral}

Several models are suggested by the previous discussion.

\noi
1.  statistical mechanics of one-dimensional systems
such as the Ising model, are described by the partition
function
\[
               Z=\sum_{\{m\}} \cdots 
             T(m_{a},m_{b}) T(m_{c},m_{d}) T(m_{e},m_{f})
                              \cdots
\]
where the transfer matrix is given by
\[
\left(
\begin{array}{cc}
               e^{J}  & e^{-J} \\
               e^{-J} & e^{J}  \\
\end{array}
\right)
\]
which is easily diagonalzed.\\
\noi
2. statistical mechanics of a disordered one-dimensional system,
where the partition function is similar to the previous example,
but the $J$ coupling is a random variable, which would correspond to
a different eigenvalue $\lambda_{i}$ at each site.\\
\noi
3. The properties of the root matrices are suggestive of the
propagator in the Dirac-Feynman path integral formulation
of quantum mechanics [12]. That is, for a 0-dimensional, two-state
system, the exchange
\[
        \left(
        \begin{array}{c}
              0\\
              1\\
         \end{array}
        \right)=
        \left(
        \begin{array}{cc}
              0&1\\
              1&0\\
         \end{array}
        \right)
        \left(
        \begin{array}{c}
              1\\
              0\\
         \end{array}
        \right)
\]
becomes
\[
        \left(
        \begin{array}{c}
              a\\
              b\\
         \end{array}
        \right)=
        \left(
        \begin{array}{cc}
              0&1\\
              1&0\\
         \end{array}
        \right)^{\frac{1}{2}}
        \left(
        \begin{array}{c}
              1\\
              0\\
         \end{array}
        \right)=
        \left(
        \begin{array}{cc}
       0.3536 + i0.8536  & -0.3536 + i0.1464\\
       -0.3536 + 0.1464  &  0.3536 + i0.8536\\
         \end{array}
        \right)
        \left(
        \begin{array}{c}
              1\\
              0\\
         \end{array}
        \right)
\]
so that there exists an intermediate state that is a superposition
of the two pure states. 
This suggests a correspondence
\[
       e^{\frac{\hbar^2}{2m}\frac{\partial^2}{\partial x^2}}
                    \rightarrow \Lambda^{\frac{1}{Y}}
\]
that is,
\[
                 \hbar^2 \sim \frac{1}{Y}
\]
To make the analogy closer,
a complete description of this process
would require a sum over all the appropriate roots;
recall that the fractional roots of the permutation matrices
results in manipulation of the eigenvalues
\[
           \lm = exp[ \frac{i2\pi}{N}]
\]
and in general the $Y^{th}$ root of the eigenvalue is obtained
from the phase as
\[
              \phi'=\frac{\phi}{Y}+\frac{i2\pi}{Y}n
\]
where
$n=1,2,...,Y$
is the multiplicity of the roots.
 A succession of the $C_{n}^{\frac{1}{Y}}$
each with different roots has the form of a path integral if
the summation over the different roots is performed,
suggestive of the ZBW effect in quantum mechanics.
This
can be written in the usual form
\[
                Z=\int D[L(x)]exp[-\int \mathfrak{L}]
\]
where $L(x)$ is related to the root of the $x^{th}$ $C$.
$\mathfrak{L}$ is effectively the log of $C$ and
path-ordering is implied.
In this expression, the integral/sum is over all paths in configuration
space for a fixed set of roots. To further complete the evaluation
a sum over all possible roots is necessary; this is easily accomplished
by working in the diagonal representation, the exponent before sum
becomes
\[
\left[
\begin{array}{cc}
                exp[-\sum_{i}\Lambda_{1}(n_{i})] &    0    \\
                 0       &  exp[-\sum_{i}\Lambda_{2}(n_{i})]   \\
\end{array}
\right]
\]
where for the $Y^{th}$ root
\[  \Lambda_{J}(n_{i})=J\frac{i2\pi}{NY} + \frac{i2\pi}{Y}n_{i}
\]
and $n_{i}=1,2,3,...,Y$. Since the exponents are all proportional,
the evaluation amounts to the need to obtain an expression for the
sum $n_{1}+n_{2}+n_{3}+...+n_{J}=M$ subject to the implied restraint.
This has the form of the central limit theorem
\[
      P(M)=\int dx_{1}p(x_{1})\int dx_{2}p(x_{2})...
              \delta(x_{1}+x_{2}+x_{3}+...+x_{J}-M)
\]
which yields a gaussian in the summation restraint variable $M$.

The resulting evaluation over a single possible configuration is
shown in Fig B. In the following figure the complex amplitudes
of the two states are displayed in the complex plane: 
that for
$|10>$ fixes the lower left corner;
that for
$|01>$ fixes the upper right corner.
The aspect ratio determines the brightness.

However, in order to make this a non-trivial problem, a non-commuting
operator is needed, such as
\[       
         \left(
         \begin{array}{ccc}
                 0  &  1  &  0  \\
                 1  &  0  &  0  \\
                 0  &  0  &  1  \\
          \end{array}
           \right).
\]
A better way to see this analogy is in the case of translation/diffusion
(i.e. classical/quantum) in one-dimension [15].
\begin{figure}[!h]
\begin{center}
\includegraphics[width=5in,height=4in]{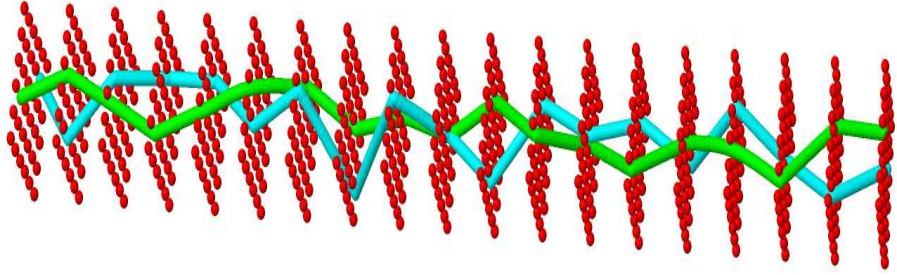}
\caption*{path of summation for two configurations}
\end{center}
\end{figure}

\begin{figure}[!h]
\begin{center}
\includegraphics[width=3in,height=3in]{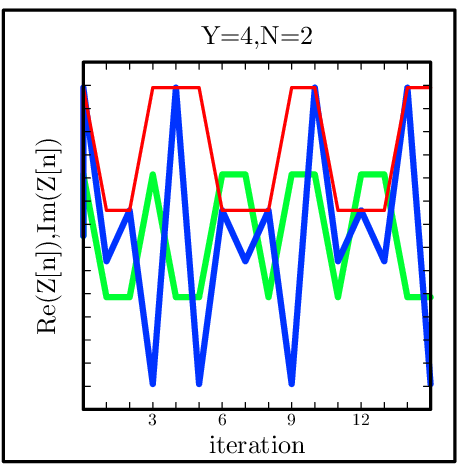}
\includegraphics[width=3in,height=3in]{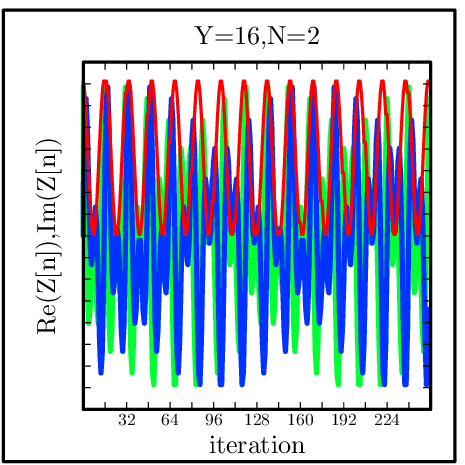}
\caption*{Fig. B: Y=4 and Y=16}
\end{center}
\end{figure}
\begin{figure}[!h]
\begin{center}
\includegraphics[width=3in,height=3in]{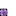}
\caption*{complex coefficients of $|10>$ and $|01>$ mapped to a rectangle}
\end{center}
\end{figure}
Increasing $Y$ is like increasing temperature.
\vspace*{27in}
\np
\noi
{\bf Discussion}

As stated in the introduction, the purpose of this note
is to outline a geometric formulation of knot descriptions
that has sufficient depth so that different knots can
be distinguished unambiguously. The simplest knots, 
3.1, 5.1, 7.1 ... represent the equivalent of Euclidean
space, the less simple knots the equivalent of curved
spaces, each distinctly described by that curvature.
Touching on the interesting, but not immediate subject of
the {\em differential permutation group} we made use of the
idea of fractional permutation to determine the normal
to the knot at a given point, and then suggested but was
not able to carry out, the use of a quantum field theory
to characterize the geometry of the knot.

The physical nature of the multiple roots is similar to that
of the anyons used in quantum computation, that is, that an
incomplete permutation is performed that creates a phase
shift in the system.
\np
\noi
{\bf Appendix:} This is not the place, nor the author, to be giving
a detailed description of the group properties of the permutation roots,
however, the subject is too interesting to leave without further
mention. Consider the case of the 3x3 cycle:
calculating the square root and then constructing the multiplication
table yields a fractal-like pattern. To see this
define
\[      \mathfrak{z}=
\left(
\begin{array}{ccc}
              0 & 1 & 0 \\
              0 & 0 & 1 \\
              1 & 0 & 0 \\
\end{array}
\right)
\]
then the square root of this in diagonal form is
\[
        \mathfrak{z}^{\frac{1}{2}}\rightarrow
\left(
\begin{array}{ccc}
       e^{\phi}  &  0  &  0  \\
         0        & e^{2\phi}  &  0  \\
         0        &   0   &  e^{3\phi} \\
\end{array}
\right)
\]
where $\phi=\frac{i2\pi}{YN}$. Initially there are $Y^{N}=2^{3}$ roots, however
if they are multiplied out, additional elements of the group are generated.
(In our notation, $Y=2,3,4,...$ for the square, cubic, quartic,... roots
of the NxN permutation matrices.)
Numbering the elements 1-24 (only the first eight are roots),
 the multiplication table is:
\tiny
\begin{center}
\[
\left[
\begin{array}{cccccccc|cccccccc|cccccccc}
 9 &10 &11 &12 &13 &14 &15 &16 &17 &18 &19 &20 &21 &22 &23 &24 & 2 & 1 & 4 & 3 & 6 & 5 & 8 & 7  \\
10 & 9 &12 &11 &14 &13 &16 &15 &18 &17 &20 &19 &22 &21 &24 &23 & 1 & 2 & 3 & 4 & 5 & 6 & 7 & 8  \\
11 &12 & 9 &10 &15 &16 &13 &14 &19 &20 &17 &18 &23 &24 &21 &22 & 4 & 3 & 2 & 1 & 8 & 7 & 6 & 5  \\
12 &11 &10 & 9 &16 &15 &14 &13 &20 &19 &18 &17 &24 &23 &22 &21 & 3 & 4 & 1 & 2 & 7 & 8 & 5 & 6  \\
13 &14 &15 &16 & 9 &10 &11 &12 &21 &22 &23 &24 &17 &18 &19 &20 & 6 & 5 & 8 & 7 & 2 & 1 & 4 & 3  \\
14 &13 &16 &15 &10 & 9 &12 &11 &22 &21 &24 &23 &18 &17 &20 &19 & 5 & 6 & 7 & 8 & 1 & 2 & 3 & 4  \\
15 &16 &13 &14 &11 &12 & 9 &10 &23 &24 &21 &22 &19 &20 &17 &18 & 8 & 7 & 6 & 5 & 4 & 3 & 2 & 1  \\
16 &15 &14 &13 &12 &11 &10 & 9 &24 &23 &22 &21 &20 &19 &18 &17 & 7 & 8 & 5 & 6 & 3 & 4 & 1 & 2  \\ \hline
17 &18 &19 &20 &21 &22 &23 &24 & 2 & 1 & 4 & 3 & 6 & 5 & 8 & 7 &10 & 9 &12 &11 &14 &13 &16 &15  \\
18 &17 &20 &19 &22 &21 &24 &23 & 1 & 2 & 3 & 4 & 5 & 6 & 7 & 8 & 9 &10 &11 &12 &13 &14 &15 &16  \\
19 &20 &17 &18 &23 &24 &21 &22 & 4 & 3 & 2 & 1 & 8 & 7 & 6 & 5 &12 &11 &10 & 9 &16 &15 &14 &13  \\
20 &19 &18 &17 &24 &23 &22 &21 & 3 & 4 & 1 & 2 & 7 & 8 & 5 & 6 &11 &12 & 9 &10 &15 &16 &13 &14  \\
21 &22 &23 &24 &17 &18 &19 &20 & 6 & 5 & 8 & 7 & 2 & 1 & 4 & 3 &14 &13 &16 &15 &10 & 9 &12 &11  \\
22 &21 &24 &23 &18 &17 &20 &19 & 5 & 6 & 7 & 8 & 1 & 2 & 3 & 4 &13 &14 &15 &16 & 9 &10 &11 &12  \\
23 &24 &21 &22 &19 &20 &17 &18 & 8 & 7 & 6 & 5 & 4 & 3 & 2 & 1 &16 &15 &14 &13 &12 &11 &10 & 9  \\
24 &23 &22 &21 &20 &19 &18 &17 & 7 & 8 & 5 & 6 & 3 & 4 & 1 & 2 &15 &16 &13 &14 &11 &12 & 9 &10  \\\hline
 2 & 1 & 4 & 3 & 6 & 5 & 8 & 7 &10 & 9 &12 &11 &14 &13 &16 &15 &18 &17 &20 &19 &22 &21 &24 &23  \\
 1 & 2 & 3 & 4 & 5 & 6 & 7 & 8 & 9 &10 &11 &12 &13 &14 &15 &16 &17 &18 &19 &20 &21 &22 &23 &24  \\
 4 & 3 & 2 & 1 & 8 & 7 & 6 & 5 &12 &11 &10 & 9 &16 &15 &14 &13 &20 &19 &18 &17 &24 &23 &22 &21  \\
 3 & 4 & 1 & 2 & 7 & 8 & 5 & 6 &11 &12 & 9 &10 &15 &16 &13 &14 &19 &20 &17 &18 &23 &24 &21 &22  \\
 6 & 5 & 8 & 7 & 2 & 1 & 4 & 3 &14 &13 &16 &15 &10 & 9 &12 &11 &22 &21 &24 &23 &18 &17 &20 &19  \\
 5 & 6 & 7 & 8 & 1 & 2 & 3 & 4 &13 &14 &15 &16 & 9 &10 &11 &12 &21 &22 &23 &24 &17 &18 &19 &20  \\
 8 & 7 & 6 & 5 & 4 & 3 & 2 & 1 &16 &15 &14 &13 &12 &11 &10 & 9 &24 &23 &22 &21 &20 &19 &18 &17  \\
 7 & 8 & 5 & 6 & 3 & 4 & 1 & 2 &15 &16 &13 &14 &11 &12 & 9 &10 &23 &24 &21 &22 &19 &20 &17 &18  \\
\end{array}
\right]
\]
\end{center}
\ns
but a better way to get a rough notion of the multiplicative structure is through
a color coding as shown in Fig C. What is of particular interest is that the
 off-diagonal elements exhibit a pattern similar to that of the cycle summation
described above.

\begin{figure}[!h]
\includegraphics[width=3in,height=3in]{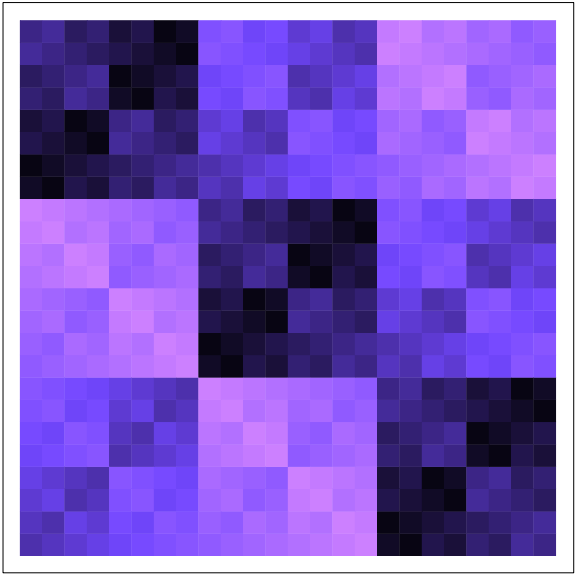}
\caption*{Fig C: $Y^{N}=2^{3}$ product table}
\end{figure}
%\begin{figure}[!h]
%\begin{center}
%\includegraphics[width=3in,height=3in]{pstsch.3.eps}
%\caption*{$Y^{N}=4^{4}$ product table}
%\end{center}
%\end{figure}

To be more specific, write the table as
\[
\left(
\begin{array}{c|c|c}
                A  &            B     &               E  \\   \hline
                B  &            E     &     \mathfrak{T}A  \\ \hline
                E  &  \mathfrak{T}A   &     \mathfrak{T}B  \\
\end{array}
\right)
\]
where
\[    \mathfrak{T}=  \ts \oplus \ts  \oplus \ts  \cdots  \]
\begin{sagesilent}
q=load('q')
r=load('r')
s=load('s')
t=load('t')
u=t*q-s
az=load('a')
m=load('mm')
\end{sagesilent}
defining the transformations that simplify $Q$ as $\mathbb{U}$
\tiny
\[
  \mathfrak{T}A=\sage{q} \]
\[  \mathfrak{T}B=\sage{r} \]
\[   \mathfrak{T}A-B=\sage{u}\]
\[   \mathbb{U}^{q}\sage{s}=\sage{az} \]\ns
which is obtained from
\[    \mathcal{F}=
       \left(
      \begin{array}{cc}
           \mathfrak{A}            & \mathcal{D}\mathfrak{A}\\
           \mathcal{D}\mathfrak{A} &            \mathfrak{A}\\
       \end{array}
        \right)
\]
and
\[     \mathfrak{A}=\sage{m} \]
and the displacement operator is
\[    \mathcal{D}=exp[-\bar{3}\partial]\]
in other words, the fractal-like appearance is explainable, at least in
this simple case, in terms of a series of matrix operations on a pair
of vectors $(1,1,1,1)$ and $(1,-1,-1,1)$.

In the above we have considered only real fractional powers, but it
is also of interest to treat the case where the power is a complex
number, e.g.
               $S_{1}^{\frac{1}{Y}}$ where $Y=Y_{r}+iY_{i}$
In this instance however, the finite nature of the group properties is lost;
if the eigenvalues acquire a "non-unitary" part, i.e. $|\lambda|\neq 1$
then the closed nature of the group table is lost.
\np
\noi
{\bf Acknowledgements:} I want to thank Michael Winocur and the UW Physics
Department for a research fellowship during the past year. In addition,
I want to thank the people of the Physics Computing Staff,
Chad Seys and Jes Tikalsky,
for their patient assistance on numerous occasions.
 I should also give special thanks to the many and varied Open Source
software organizations online.
\\
\\
\\
\noi
{\bf Software:}\\
\noi
{\em Computation:} Most of the numerical work indicated here was done using
Linux based, home spun C++ coding,
then later checked using  either the {\em SAGEmath} package [13a],
or {\em MATHEMATICA} [13b].
As far as I could tell neither {\em SAGE} nor {\em MATHEMATICA}
provided direct access to the
multiple roots of the permutation operators,
but both were easily coerced to work with the eigenvalues.
\\
\noi
{\em Graphics:} The outlines of the 5.2 and 9.2 knots were made
initially using the KnotTheory package [7],
then edited using {\em POSTSCRIPT} commands;
the 3.1 and 7.2 knots, as well as Fig B, were constructed
using the linear algebra facilities available in {\em METAPOST};
Fig. A and the matrix plots in the Appendix
were drawn using a special {\em METAPOST} plotting
module [14];
the lattice figure was constructed using the 3D plotting 
modules in {\em SAGE}.
This article was written using {\em SAGETEX} [10a], a typesetting miracle.
\\
\\
\\
\\
{\bf References:}\\

\noi
[1] V. Manturov, {\em Knot Theory}, Chapman and Hall (2004)\\
\noi
[2] J. Hoste, M. Thistlewaite, J. Weeks, {\em The First 1.7 Million Knots}
Mathematical Intellegencier \underline{20} 33 (1998)\\
\noi
[3] J. Pachos, {\em Intro to Topological Quantum Computing}, Cambridge (2012)\\
\noi
[4] J. Preskill, {\em Lectures on Quantum Computation}, Cal Tech Physics 229(219) (1997-2004)\\
\noi
[5] L. Kauffman, {\em Physics and Knots}, World Scientific (1991)\\
\noi
[6] Perko Pair: A brief introduction to this very interesting piece
of history is found at Wikipedia
http://en.wikipedia.org/wiki/Perko\_Pair\\
\noi
[7]
These drawings were made initially with the {\em KnotTheory} package obtained from:\\
 http://katlas.math.toronto.edu.\\
\noi
[8] http://katlas.org/wiki/Heegaard\_Floer\_Knot\_Homology\\
\noi
[9a] Most graphics packages today have some facilities in linear algebra,
but few provide the full extent of these properties; however, newer graphics programs
implement the Clifford algebra - commonly known as Geometric algebra - formalism
which better emphasizes the additive nature of visual images;\\
\noi
[9b] on another issue,
object-oriented design used in nearly all modern
computer projects may be a powerful tool for the programmer, but is a
distraction in use; the need to be able add a line, a disk and a
cube, to an existing frame is a geometric demand, easily carried out in
the context of a graded algebra, regardless of the internal coding;
See, for a differing view and additional references,
D. Hildenbrand , D. Fontijne , C. Perwass and L. Dorst,
{\em Geometric Algebra and its Application to Computer Graphics}
http://www.science.uva.nl/ga/ \\
\noi
[10] L. Dorst et al, {\em Geometric Algebra and Computer Science} Morgan Kaufmann
  (2011); C Doran, A. Lasby {\em Geometric Algebra for Physicists} Cambridge
(2003)\\
\noi
[11] R. Penrose, in {\em Quantum Theory and Beyond}, T Bastin ed. Cambridge (1971)\\
\noi
[12] RP Feynman, AR Hibbs; {\em Path Integrals and QM} McGraw-Hill (1964); 
A. Zee, {\em QFT in a Nutshell}, Princeton (2004)\\
\noi
[13a]
 William A. Stein et al., Sage Mathematics Software (Version 5.8).
The Sage Development Team, March 2013, http://www.sagemath.org.\\
\noi
[13b] Wolfram Mathematica 9.0.1.0;  http://www.wolfram.com\\
\noi
[14] R. Kariotis, "Plotting Data With Metapost", unpublished but
supposedly online somewhere\\
\noi
[15] R. Kariotis, "Translation/Diffusion in Complex Space", to be published\\
\end{document}